\documentclass[aps,apl,onecolumn,superscriptaddress,amsmath,amssymb,showpacs,amsfronts]{revtex4}
\usepackage{color}
\usepackage{graphics}

\begin{document}

\title{On the Computation and Applications of Bessel Functions with Pure Imaginary Indices}

\author{A.~A.~Matyshev}
\affiliation{Department of Physical Electronics, St Petersburg State Technical University, Politekhnitcheskaya-Str 29, 195251 St Petersburg, Russia}

\author{E.~Fohtung}
\email[]{edwin.fohtung@iss.fzk.de}
\affiliation{Institute for Synchrotron radiation, Forschungszentrum Karlsruhe, 76344 Eggenstein-Leopoldshafen, Germany}


\begin{abstract}
\noindent Bessel functions with pure imaginary index (order) play an important role in corpuscular optics where they govern the
dynamics of charged particles in isotrajectory quadrupoles. Recently they were found to be of great importance in
semiconductor material characterization as they are manifested in
the strain state of crystalline material. A new algorithm which
can be used for the computation of the normal and modified Bessel
functions  with pure imaginary index is proposed. The developed
algorithm is very fast to compute and for small arguments
converges after a few iterations.

\end{abstract}


\maketitle
\section{Introduction}

\noindent Bessel functions occur in many branches of mathematical physics as
solutions of differential equations when boundary conditions such as
the Dirichlet or Neumann are imposed on various space domains. The Bessel functions with index
(order) $\nu$ can be represented as the solutions of the following
differential equation \cite{Watson45}:

\begin{equation}
\label{1}
{x^2}y'' + xy' + \left( {{x^2} - {v^2}} \right)y = 0,
\end{equation}

\noindent where $y'$ and $y''$ are the first and second derivatives (respectively) with respect to $x$ while $\nu$
is a complex constant. The solutions of eqn.(\ref{1}) can be expressed as an absolute converging series that is defined in the entire complex plane:

\begin{eqnarray}
\label{2}
	{J_v}\left( z \right) = {\left( {\frac{x}{2}} \right)^v}{\sum\limits_{n = 0}^\infty  {\left( { - 1} \right)} ^n}\frac{1}{{n!\Gamma \left( {v + n + 1} \right)}}{\left( {\frac{x}{2}} \right)^{2n}}.
\end{eqnarray}

\noindent In a great variety of applications such as diffraction, electrical induction, etc, only Bessel functions of the zeroth and first
orders, namely, $J_0(x)$  and $J_1(x)$  occur while in other physical applications (such as solutions of Kepler's equation), the entire
indices  $n$  are used. Bessel functions with complex indices were considered to have limited areas of applicability in natural and applied sciences till recently.

\noindent The normal and modified Bessel function with pure imaginary index was shown to be the solution governing the motion of charged particles in isotrajectory quadrupoles \cite{Matyshev00}. Recently these functions have been found to be present as solutions of the Lame's equation which characterizes the displacement (and strain) field distribution in semiconductor nanostructures \cite{Fohtung09}.

\noindent In order to compute the solutions of equation (\ref{1}) for the case of purely imaginary indices $i\nu$, it is required to obtain the sum of a series with each coefficient being a complex number and  to compute the complex value function $\Gamma(i\nu)$ which on its own is a very laborious operation. In reality, for the case of a purely imaginary index $i\nu$ and for a natural number $n$ we have:

\begin{eqnarray}
\label{3}
\Gamma(i\nu+n+1)=\Gamma(i\nu)\,\prod\limits_{m=0}^n (i\nu+m)
\end{eqnarray}
and thereafter Bessel function (\ref{2}) may be re-expressed in the form:
\begin{equation}\label{4}
J_{i\nu}(x)=\frac{1}{\Gamma(i\nu)}{\left(\frac{x}{2}\right)}^{i\nu}
\sum\limits_{n=0}^{\infty} {(-1)}^n
\frac{1}{n!\,\prod\limits_{m=0}^n (i\nu+m)}
{\left(\frac{x}{2}\right)}^{2n}\,,
\end{equation}

\noindent where $\nu$ and $x$ are real and complex numbers respectively. The solution of (\ref{1}) expressed in the terms of (\ref{4}) represents a "difficult-to-compute" complex solution of a differential equation with the real coefficients. To the best of our knowledge there is no concrete algorithm that can enable us to compute Bessel functions with purely imaginary order as they have had no applications in areas of natural sciences till date. Only a hand full of pure mathematicians made attempts to investigate such functions.

\noindent The author of one of the most detailed treatment on Bessel functions \cite{Watson45} thought that these functions were of no interest though they were noticed by Lommel \cite{Lommel1871} who defined the function $J_{\nu+i\mu}(x)$ in the form

\begin{equation}\label{5}
  J_{\nu+i\mu}(x)=\frac{{(x/2)}^{\nu+i\mu}}{\Gamma({\nu+i\mu}+1/2)\Gamma(1/2)}
\left[K_{\nu,\,\mu}(x)+i\,S_{\nu,\,\mu}(x)\right]
\end{equation}
\noindent with $K$ and $S$ being real valued functions. Lommel was motivated by the following differential equation:
\begin{equation}\label{6}
 x^2y''+(2\alpha-2\beta\nu+1)xy'+\left[\alpha(\alpha-2\beta\nu)+
{\beta}^2{\gamma}^2x^{2\beta}\right]y=0\,.
\end{equation}
If any real $\beta\ne 0$ then equation (\ref{6}) has a solution
\begin{equation}\label{7}
  y=x^{\beta\nu-\alpha}\left[AJ_{\nu}(\gamma x^{\beta})+BJ_{-\nu}(\gamma x^{\beta}) \right]\,.
\end{equation}

\noindent Comparing equation (\ref{7}) with the equation
\begin{equation}\label{800}
x^2y''+axy'+(b+cx^{2\beta})y=0\,,
\end{equation}
\noindent Lommel found a relationship between the coefficients of equation (\ref{7}) and those of equation (\ref{8}) and expressed them in the form:
 \begin{eqnarray}
\label{8}
 \beta \nu - \alpha  =  - {{\left( {a - 1} \right)} \mathord{\left/
 {\vphantom {{\left( {a - 1} \right)} 2}} \right.
 \kern-\nulldelimiterspace} 2}, \\
\label{9}
 \beta \nu = \sqrt {{{\left[ {{{\left( {a - 1} \right)} \mathord{\left/
 {\vphantom {{\left( {a - 1} \right)} 2}} \right.
 \kern-\nulldelimiterspace} 2}} \right]}^2} - b} ,{\rm{  }}\beta \gamma = \sqrt c.
\end{eqnarray}

\noindent It turned out that the real valued coefficients $\left({a,b,c} \right)$ produced in a general case solution of equation (\ref{6}) with a complex index $\nu$. At the end Lommel obtained a very cumbersome solution of equation (\ref{6}) as multiples of a complex valued functions. He did not propose a method for the calculation of the real valued functions $K$ and  $S$ from (\ref{5}).

\noindent Another mathematician Bocher encountered the modified Bessel function with pure imaginary index while he ponderered over the solutions of Laplace's equation using the method of variable separation in cylindrical coordinate system \cite{Bocher1892}. As opposed to Lommel \cite{Lommel1871}, Bocher proposed real valued solutions with the aid of real valued series without studying the convergence of these series. Each of the term of Bocher's series was found to be a ratio of polynomial functions in $\nu$, with coefficients that increased n the order $n!$.This possed a major problem for the calculation and computation via his representation.

\noindent Finally the McDonald's function ${K_{i\tau }}\left( x \right)$ also having pure imaginary index was extensively used in its integral
form to obtain solutions of the Laplace and wave equations for different boundary value conditions by the Soviet mathematicians M.
Kantorovitch and N. Lebedev \cite{Greenberg48}. They used the integral representation

\begin{eqnarray}
\label{10} {K_{i\tau }}\left( x \right) = \int\limits_0^\infty
{\exp \left( { - x\cosh t} \right)\cos \left( {\tau t} \right)}
dt,{\rm{  }}x > 0.
\end{eqnarray}

\noindent This representation are convenient only for sufficiently large values of the argument $x$. Unfortunately, for small values of $x$ this integral representation is not quite effective, and this is the case where Bessel functions with pure imaginary indices become applicable and useful in charge particle dynamics \cite{Matyshev00} and strain field investigations in nanostructures \cite{Fohtung09}.

The above summary shows that there is no  effective algorithm for the computation of Bessel functions with pure imaginary indices. In the next sections, we provide this algorithm.

\section{Method and Calculations}

\noindent So, Bessel functions with pure imaginary indices are solutions of the equation
\begin{equation}\label{100}
   x^2y''+xy'+(x^2+{\nu}^2)y=0\,,
\end{equation}
where $\nu$ is a real number. The pure mathematician George Boole was not interested in Bessel functions but was the first person who, over a century and a half ago developed an operational method for solving differential equations. Nevertheless in a small paragraph of his article \cite{Boole1844}, equation (\ref{100}) was mentioned and the substitution that facilitated the solving of the problem of calculating  Bessel functions with purely imaginary order was used.
So expression (\ref{2}) suggests the form of the solutions of equation (\ref{100}):

 \begin{eqnarray}
\label{11}
 y = A\left( x \right)\cos \left( {\nu \ln x} \right) + B\left( x \right)\sin \left( {\nu\ln x}
 \right)\,,
\end{eqnarray}

\noindent where $A\left( x \right)$ and $B\left( x \right)$ are series with the coefficients as shown:

 \begin{eqnarray}
\label{12}
 \begin{array}{l}
 A\left( x \right) = {a_0} + {a_1}x + {a_2}{x^2} +  \cdot  \cdot  \cdot  + {a_n}{x^n} +  \cdot  \cdot  \cdot , \\
 B\left( x \right) = {b_0} + {b_1}x + {b_2}{x^2} +  \cdot  \cdot  \cdot  + {b_n}{x^n} +  \cdot  \cdot  \cdot . \\
 \end{array}
\end{eqnarray}

\noindent Substituting equation (\ref{11}) into equation (\ref{100}) and equating the coefficients before the linearly
independent functions $\cos \left( {\nu\ln x} \right){\rm{ and
}}\sin \left( {\nu\ln x} \right)$ to zeros, firstly we obtain the
relation
 \begin{eqnarray}
\label{13}
 {a_1} = {b_1} = 0,
\end{eqnarray}
\noindent and secondly $\forall n \ge 2$ the recurrent relationships

\begin{eqnarray}
\label{14}
 {a_n} =  - \frac{{n{a_{n - 2}} - 2\nu{b_{n - 2}}}}{{n\left( {{n^2} + 4{\nu^2}} \right)}}, \\
\label{15} {b_n} =  - \frac{{ - 2\nu{a_{n - 2}} + n{b_{n -
2}}}}{{n\left( {{n^2} + 4{\nu^2}} \right)}}\,.
\end{eqnarray}
\noindent It is clearly seen from equations (\ref{14}) and (\ref{15})  that odd term coefficients take on zero values while those with even terms can be obtained via a recurrent relations using the values ${a_0},{\rm{ }}{b_0}$. The solution of the Bessel equation (\ref{100}) in the form (\ref{11}) was firstly introduced by Boole in \cite{Boole1844}. However, he did not analyze the convergence of the series (\ref{12}). Boole's recurrent relationship was completely forgotten although they can be easily simplified and used to effectively calculate the functions under examination. Since  only coefficients with even indices occur, we can re-define the functions $A\left( x \right)$ and  $B\left( x \right)$  in the form:

\begin{eqnarray}
\label{16} A\left( x \right) = \sum\limits_{n = 0}^\infty
{{a_{2n}}} {2^{2n}}{\left( {\frac{x}{2}} \right)^{2n}} =
\sum\limits_{n = 0}^\infty  {{A_{2n}}} {\left( {\frac{x}{2}}
\right)^{2n}}, \\ \label{17}
 B\left( x \right) = \sum\limits_{n = 0}^\infty  {{b_{2n}}} {2^{2n}}{\left( {\frac{x}{2}} \right)^{2n}} = \sum\limits_{n = 0}^\infty  {{B_{2n}}} {\left( {\frac{x}{2}} \right)^{2n}}.
\end{eqnarray}
\noindent Now for $n \ge 1$ the recurrent relationship for the coefficients ${A_{2n}},{\rm{ }}{{\rm{B}}_{2n}}$ can be simplified:

 \begin{eqnarray}
\label{18} {A_{2n}} =  - \frac{{n{A_{2n - 2}} -\nu{B_{2n -
2}}}}{{n\left( {{n^2} + {\nu^2}} \right)}},\\ \label{19}
 {B_{2n}} =  - \frac{{  \nu{A_{2n - 2}} + n{B_{2n - 2}}}}{{n\left( {{n^2} + {\nu^2}} \right)}}.
\end{eqnarray}

\section{Construction of Computable Solutions and proof of Convergence}

\noindent In order to investigate and numerically compute the strain field distribution in nanostructures, or to numerically implement these Bessel
functions in other areas of physics, we need to be able to demonstrate that the solutions to this differential equation exist and above of all converges in our domain of definition and interest. We now show that the series (\ref{14}), (\ref{15}) converges and does so absolutely for any complex valued argument $x$ and order $v$. As a series will converge absolutely if and only if the absolute value of the ${n^{th}}$ term (which we shall refer to as the majorant) converges. Let the majorant ${M_{2n}}$ be defined such that

\begin{eqnarray}
\label{20}
 {M_{2n}} = \left| {{A_{2n}}} \right| + \left| {{B_{2n}}}
 \right|\,.
\end{eqnarray}

\noindent It follows from (\ref{18}), (\ref{19}), (\ref{20}) that

\begin{multline}
\label{21}
	{M_{2n}} \le \left[ {\frac{1}{{{n^2} + {v^2}}} + \frac{{\left| v \right|}}{{n\left( {{n^2} + {v^2}} \right)}}} \right]{M_{2n - 2}} \\ \le \left( {\frac{1}{{{n^2}}} + \frac{{\left| v \right|}}{{{n^3}}}} \right){M_{2n - 2}}.
\end{multline}

\noindent The last inequality provides us the possibility of studying the order of which the majorant ${M_{2n}}$ decays (or grows). It can be easily shown that

\begin{eqnarray}
\label{22}
 {M_{2n}} \le C\frac{{{n^{\left|\nu \right|}}}}{{\left( {n!} \right)^2}},
\end{eqnarray}
\noindent where $C$ is a positive constant dependent on the zero-th terms of the sequence. Using the d'Alembert's test for convergence the
proof of absolute convergence for all values of $x$ and $\nu$, is completed.

\noindent Now it is possible to form, without lost of generality, two linearly independent solutions of (\ref{100}). Let us choose two pairs of
the values for ${A_0},{\rm{ }}{B_0}$ that generates the sequences ${A_{2n}},{\rm{ }}{B_{2n}}$ and its corresponding functions (\ref{16}) and (\ref{17}). For brevity, these two pairs provide two solutions (\ref{100}) that can be easily computed numerically for a few number of iterations. In \cite{Matyshev00}, it was shown that the solution spawned from the pair

	\begin{eqnarray}
\label{23}
 \left( {{A_0},{B_0}} \right) = \left( {0,1} \right),
\end{eqnarray}
\noindent  represented as $\mathrm{Sf}_{\nu}(x)$ and the ones that are spawned from the pair

\begin{eqnarray}
\label{24}
 \left( {{A_0},{B_0}} \right) = \left( {1,0} \right),
\end{eqnarray}
\noindent was also represented as $\mathrm{Cf}_{\nu}(x)$. Then an analytical expression of these functions can be easily obtained:

\begin{multline}
\label{25} \mathrm{Sf}_{\nu}(x) = \left[ {1 - \frac{1}{{1 +
{\nu^2}}}{{\left( {\frac{x}{2}} \right)}^2} +  \cdots  }
\right]\sin \left( {\nu\ln x} \right) \\ + \left[ {\frac{\nu}{{1 +
{\nu^2}}}{{\left( {\frac{x}{2}} \right)}^2} +  \cdots} \right]\cos
\left( {\nu\ln x} \right),
\end{multline}
\begin{multline}
 \label{26}
 \mathrm{Cf}_{\nu}(x) = \left[ {1 - \frac{1}{{1 + {\nu^2}}}{{\left(
{\frac{x}{2}} \right)}^2} + \cdots} \right]\cos \left( {\nu\ln x}
\right) \\ + \left[ {\frac{\nu}{{1 + {\nu^2}}}{{\left(
{\frac{x}{2}} \right)}^2} + \cdots} \right]\sin \left( {\nu\ln x}
\right).
 \end{multline}

\noindent For $x \to 0$ we have two equivalent relationships

\begin{eqnarray}
\label{27} \mathrm{Sf}_{\nu}(x) \sim \sin \left( {\nu\ln x}
\right)
 , \\
 \label{28}
\mathrm{Cf}_{\nu}(x) \sim \cos \left( {\nu\ln x} \right).
\end{eqnarray}

\noindent Now the general solution of the differential equation (\ref{100}) with real coefficients may be explicitly written out via the real
valued functions

\begin{eqnarray}
\label{29} y = {c_1}\mathrm{Sf}_{\nu}(x) +
{c_2}\mathrm{Cf}_{\nu}(x)\,,
\end{eqnarray}
where $c_1$ and $c_2$ are any real constants. Provided that equation (\ref{100}) has a complex valued solution in the form of the Bessel function (\ref{4}). Therefore the latter must be the linear combination of the functions $\mathrm{Sf}_{\nu}(x)$ and $\mathrm{Cf}_{\nu}(x)$. It is easy to
show that

\begin{equation}\label{300}
J_{i\nu}(x)=\mathrm{Cf}_{\nu}(x)+i\mathrm{Sf}_{\nu}(x)\,.
\end{equation}
In other words,the functions $\mathrm{Sf}_{\nu}(x)$ and $\mathrm{Cf}_{\nu}(x)$ are the real and imaginary parts of Bessel function (\ref{4}). It follows from the following equivalent relationships:

$$J_{i\nu}(x)=\frac{1}{\Gamma(i\nu+1)}\left(\frac{x}{2}\right)^{i\nu}
\left[1-\frac{1}{i\nu+1}\left(\frac{x}{2}\right)^2+\cdots\right]\sim
$$ $$\sim\frac{1}{i\nu\Gamma(i\nu)2^{i\nu}}\left[\cos(\nu\ln
x)+i\sin(\nu\ln x)\right]\sim \left[\cos(\nu\ln x)+i\sin(\nu\ln
x)\right]\,. $$ The first equivalent relationship is valid as long
as $x\rightarrow 0$ and the second\footnote{To prove the second
equivalent relationship it is necessary to use the identity
$$\Gamma(i\nu)\Gamma(1-i\nu)\equiv\frac{\pi}{\sin{i\nu}}\,.$$} one
is valid so long as $\nu\rightarrow 0$.

\begin{eqnarray}
\label{30} {x^2}y'' + xy' + \left( { - {x^2} + {\nu}^2} \right)y =
0.
\end{eqnarray}

\noindent As

\begin{equation}
\label{31} \cos \left( {\nu\ln ix} \right) = \cosh \left(
{\frac{{\pi \nu}}{2}} \right)\cos \left( {\nu\ln x} \right)  -
\sinh \left( {\frac{{\pi \nu}}{2}} \right)\sin \left( {\nu\ln x}
\right),
 \end{equation}

\begin{equation}
\label{32} \sin \left( {\nu\ln ix} \right) =  \cosh \left(
{\frac{{\pi \nu}}{2}} \right)\sin \left( {\nu\ln x} \right)  +
\sinh \left( {\frac{{\pi \nu}}{2}} \right)\cos \left( {\nu\ln x}
\right),
 \end{equation}

\noindent and as

\begin{eqnarray}
\label{33}
 C\left( x \right) = A\left( {ix} \right) = \sum\limits_{n = 0}^\infty  {{C_{2n}}} {\left( {\frac{x}{2}} \right)^{2n}}, \\
       \label{34}
 D\left( x \right) = B\left( {ix} \right) = \sum\limits_{n = 0}^\infty  {{D_{2n}}} {\left( {\frac{x}{2}} \right)^{2n}},
 \end{eqnarray}

\noindent it is possible to construct two linearly independent solutions for this case in the form
\begin{eqnarray}
\label{37} {y_1} = C\left( x \right)\cos \left( {\nu\ln x} \right)
+ D\left( x \right)\sin \left( {\nu\ln x} \right),\\ \label{38}
{y_2} = D\left( x \right)\cos \left( {\nu\ln x} \right) - C\left(
x \right)\sin \left( {\nu\ln x} \right),
 \end{eqnarray}

\noindent where as earlier mentioned, for $n \ge 1$ the recurrent equations for ${C_{2n}},{\rm{ }}{D_{2n}}$ takes the expected form
\begin{eqnarray}
\label{35}
 {C_{2n}} = \frac{{n{C_{2n - 2}} - \nu{D_{2n - 2}}}}{{n\left( {{n^2} + {{\nu}^2}} \right)}}, \\
\label{36}
 {D_{2n}} = \frac{{\nu{C_{2n - 2}} + n{D_{2n - 2}}}}{{n\left( {{n^2} + {{\nu}^2}} \right)}}.
 \end{eqnarray}
As the series $C(x)$ and $D(x)$ have the same majorant (\ref{22}), these series will thus converge absolutely for any $x$ and $\nu$. One real valued solution of (\ref{100}) spawns two real valued solutions for (\ref{30}). Thus by choosing only one pair of values ${C_0},{D_0}$ it is possible to obtain two real valued functions being solutions of equation (\ref{30}). Let

\begin{eqnarray}
\label{39}
 \left( {{C_0},{D_0}} \right) = \left( {0,1} \right)\,.
 \end{eqnarray}
Substituting (\ref{33}) and (\ref{34}) into  (\ref{37}) and (\ref{38}) it is possible to obtain two functions which was denoted in \cite{Matyshev00} as $\mathrm{Sd}_{\nu}\left( x \right)$ and $\mathrm{Cd}_{\nu}\left( x \right)$:

\begin{multline}
\label{40} \mathrm{Sd}_{\nu}\left( x \right) = \left[ {1 +
\frac{1}{{1 + {{\nu}^2}}}{{\left( {\frac{x}{2}} \right)}^2} +
\cdots} \right]\sin \left( {\nu\ln x} \right) \\ + \left[ { -
\frac{\nu}{{1 + {{\nu}^2}}}{{\left( {\frac{x}{2}} \right)}^2} +
\cdots } \right]\cos \left( {\nu\ln x} \right),
\end{multline}

\begin{multline}
\label{41}
 \mathrm{Cd}_{\nu}\left( x \right) = \left[ {1 + \frac{1}{{1 + {{\nu}^2}}}{{\left( {\frac{x}{2}} \right)}^2} +
  \cdots } \right]\cos \left( {\nu\ln x} \right) \\
  - \left[ { - \frac{\nu}{{1 + {{\nu}^2}}}{{\left( {\frac{x}{2}} \right)}^2} +
    \cdots } \right]\sin \left( {\nu\ln x} \right)\,.
\end{multline}

\noindent As $x \to 0$ we also have two equivalent relationships

\begin{eqnarray}
\label{42} \mathrm{Cd}_{\nu}\left( x \right) \sim \cos \left(
{v\ln x} \right), \\ \label{43}  \mathrm{Sd}_{\nu}\left( x \right)
\sim \sin \left( {v\ln x} \right)\,.
 \end{eqnarray}

\noindent It is obvious that

\begin{equation}\label{700}
  J_{i\nu}(ix)=\mathrm{Cd}_{\nu}\left( x \right)+i\mathrm{Sd}_{\nu}\left( x
  \right)\,,
\end{equation}

\noindent where $x$ and $\nu$ are any real numbers.

\section{Wronskians of functions}

\noindent It can be easily shown that the Wronskian $W$ of the two linearly independent solutions of the equation

\begin{eqnarray}
\label{44}
 y'' + \frac{1}{x}y' + f\left( x \right)y = 0
\end{eqnarray}

\noindent has the form $W = a/x$, where the constant $a$ depends on the pair of concrete solutions. From the equivalent relationships (\ref{27}), (\ref{28}) and (\ref{42}), (\ref{43}) it follows that

\begin{eqnarray}
\label{45} \mathrm{{Cf}_{\nu}}\left( x \right){\left[
\mathrm{{Sf}_{\nu}}\left( x \right) \right]^\prime } -
\mathrm{{Sf}_{\nu}}\left( x \right)\left[
\mathrm{{Cf}_{\nu}}\left( x \right) \right]^\prime  =
\frac{\nu}{x}, \\ \label{46}
 \mathrm{{Cd}_{\nu}}\left( x \right){\left[
\mathrm{{Sd}_{\nu}}\left( x \right) \right]^\prime } -
\mathrm{{Sd}_{\nu}}\left( x \right)\left[
\mathrm{{Cd}_{\nu}}\left( x \right) \right]^\prime  =
\frac{\nu}{x}\,.\end{eqnarray}

\noindent It is clearly seen that the Wronskians (\ref{45}) and (\ref{46}) vanishes for $\nu = 0$. As $\nu \to 0$ we have

\begin{eqnarray}
 \mathrm{{Cf}_{\nu}}\left( x \right) \to {J_0}\left( x \right),  \\
\mathrm{{Sf}_{\nu}}\left( x \right) \to 0
\end{eqnarray}

\noindent and
\begin{eqnarray}
 \mathrm{{Cd}_{\nu}}\left( x \right) \to {I_0}\left( x \right),  \\
\mathrm{{Sd}_{\nu}}\left( x \right) \to 0\,.
\end{eqnarray}

\section{Accuracy of proposed Computational Algorithm}

\noindent To obtain a measure of the computational error for the proposed procedure, we require a detailed study of the majorant
(remainder) ${M_{2n}}$. Let

\begin{eqnarray}
\label{47} {M_{2n}} = \frac{{{n^{\left|\nu \right|}}}}{{{{\left(
{n!} \right)}^2}}}{m_{2n}}.
\end{eqnarray}

\noindent Now it is possible to obtain an estimate of remainder ${m_{2n}}$ for $n\ge2$

\begin{eqnarray}
\label{48} \frac{{{m_{2n}}}}{{{m_{2n - 2}}}} \le \left( {1 +
\frac{{\left| \nu \right|}}{n}} \right){\left( {1 - \frac{1}{n}}
\right)^{\left| \nu \right|}}.
\end{eqnarray}

\noindent The Lagrange's formula for remainder of a Taylor's series gives for $n \ge 2$

\begin{eqnarray}
\label{49}
 \frac{{{m_{2n}}}}{{{m_{2n - 2}}}} \le 1 - \frac{{{{\nu}^2}}}{{{n^2}}} + \frac{{\left| \nu \right|
 \left( {\left| \nu \right| - 1} \right)}}{{6{n^3}}}{\left( {1 - \frac{\theta }{n}} \right)^{\left| \nu \right| - 2}}
  \times  \nonumber \\
 {\rm{       }} \times \left[ {3\left| \nu \right| + \left( {\left| \nu \right| - 2} \right)
 \left( {1 + \left| \nu \right|\frac{\theta }{n}} \right){{\left( {1 - \frac{\theta }{n}}
  \right)}^{\left| \nu \right| - 3}}} \right],
 \end{eqnarray}

\noindent where $0 \le \theta  \le 1.$  The last inequality may be improved by removing from the right hand side the quantity
$\theta$:

\begin{equation}
\frac{{{m_{2n}}}}{{{m_{2n - 2}}}} \le 1 -
\frac{{{{\nu}^2}}}{{{n^2}}} + \frac{F}{{{n^3}}}\,,
\end{equation}

\noindent where

\begin{center}

\begin{equation}
F=\left\{
\begin{array}{c}
{\nu}^2\left||\nu|-1\right|2^{(1-|\nu|)}\,\quad |\nu| \le 2\,,\\
|\nu|\left||\nu|-1\right|
\left[3|\nu|+(|\nu|-2)(1+|\nu|/2)2^{3-|\nu|}\right]/6\,,\quad 2 <
|\nu| \le 3\,,\\ |\nu|\left||\nu|-1\right|
\left({\nu}^2/2+3|\nu|-2\right)/6\,,\quad |\nu|>3\,.
\end{array}
\right.
\end{equation}
\end{center}

\noindent As for $\nu\ne0$

\begin{equation}
\label{50} \frac{{{m_{2n}}}}{{{m_{2n - 2}}}} \le 1 -
\frac{{{{\nu}^2}}}{{{n^2}}} + \frac{F}{{{n^3}}} < 1 +
\frac{{{{\nu}^2}}}{{{n^2}}} + \frac{F}{{{n^3}}},
\end{equation}

\noindent then it is possible to obtain

\begin{equation}
\ln \left( {{m_{2n}}} \right) - \ln \left( {{m_{2n - 2}}} \right)
< \ln (1 + \frac{{{{\nu}^2}}}{{{n^2}}} + \frac{F}{{{n^3}}}) <
\frac{{{{\nu}^2}}}{{{n^2}}} + \frac{F}{{{n^3}}}\,. \nonumber
\end{equation}

\noindent Using the well known sums

\begin{eqnarray}
\sum\limits_{n = 2}^\infty  {\frac{1}{{{n^2}}}}  = \frac{{{\pi
^2}}}{6} - 1 =0.6449\,,\nonumber \\ \sum\limits_{n = 2}^\infty
{\frac{1}{{{n^3}}} = \zeta (3) - 1 = 0.2021} \nonumber
\end{eqnarray}

\noindent and also the values

$$\begin{array}{l}
 {m_0} = 1\,, \\
 {m_2} = \frac{{1 + \left| \nu \right|}}{{1 + {{\nu}^2}}\,,} \\
 \end{array}$$

\noindent it is possible finally to obtain the inequality

\begin{eqnarray}
\label{51} \ln \frac{{{m_{2n}}}}{{{m_{2n - 2}}}} < 0.6449{{\nu}^2}
+ 0.2021F\,.
\end{eqnarray}

\noindent From the last inequality it is easy to obtain a final one simplified in the form

\begin{eqnarray}
{m_{2n}} < \frac{{1 + \left| \nu \right|}}{{1 + {{\nu}^2}}}\exp
(0.6449{v^2} + 0.2021F) = m(\nu)\,.
\end{eqnarray}

\noindent Now we have

\begin{eqnarray}
\label{52} {{\rm{M}}_{{\rm{2n}}}} < m(\nu)\frac{{{n^{\left| \nu
\right|}}}}{{{{(n!)}^2}}}
\end{eqnarray}

\noindent and it is possible to conclude that the sum of the remaining terms for the series of functions $A\left( x \right),B\left( x \right),C\left( x \right),D\left( x \right)$ cannot exceed the value

\begin{eqnarray}\label{530}
\varepsilon  = m(\nu)\sum\limits_{n = N}^\infty {\frac{{{n^{\left|
\nu \right|}}}}{{{{(n!)}^2}}}} {\left( {\frac{x}{2}}
\right)^{2n}}\,.
\end{eqnarray}

\noindent Since the remainder (\ref{530})  depends on the value of $\nu$, let us consider for example the case $\left| \nu \right| \le 2$. Then

\begin{multline}
\label{53} {\varepsilon _N} \le
m(\nu)\frac{{{x^2}}}{4}\sum\limits_{n = N + 1}^\infty
{\frac{1}{{{{\left[ {\left( {n - 1} \right)!} \right]}^2}}}}
{\left( {\frac{x}{2}} \right)^{2n - 2}}  = \\
m(\nu)\frac{{{x^2}}}{4}\sum\limits_{n = N}^\infty
{\frac{1}{{{{(n!)}^2}}}} {\left( {\frac{x}{2}} \right)^{2n}}\,.
\end{multline}

\noindent The sum at the right hand side of equation (\ref{53}) is simply the remainder of the series for the modified Bessel function ${I_0}\left( x \right)$. This remainder may be evaluated further as shows below:

\begin{multline}
 \sum\limits_{n = N}^\infty  {\frac{1}{{{{\left( {n!} \right)}^2}}}{{\left( {\frac{x}{2}} \right)}^{2n}} =
  \frac{1}{{{{\left( {N!} \right)}^2}}}} {\left( {\frac{x}{2}} \right)^{2N}}
  \sum\limits_{n = 0}^\infty  {\frac{{{{\left( {N!} \right)}^2}}}{{{{\left[ {\left( {n + N} \right)!} \right]}^2}}}}
   {\left( {\frac{x}{2}} \right)^{2n}} =  \\
 {\rm{   }} = \frac{1}{{{{\left( {N!} \right)}^2}}}{\left( {\frac{x}{2}} \right)^{2N}}
 \left\{ {1 + \frac{1}{{{{\left( {1 + N} \right)}^2}}}} \right.{\left( {\frac{x}{2}} \right)^2} +  \\
 {\rm{   }} \cdots  + \left. {\frac{{{{1}}}}{{{{\left[ {\left( {1 + N} \right)
 \left( {2 + N} \right) \cdots \left( {n + N} \right)} \right]}^2}}}\left(\frac{x}{2}\right)^{2n} +  \cdots } \right\} <  \\
 {\rm{  }} < \frac{1}{{{{\left( {N!} \right)}^2}}}{\left( {\frac{x}{2}} \right)^{2N}}
 \left[ {1 + \frac{1}{{1  \cdot 2}}{{\left( {\frac{x}{2}} \right)}^2} +\cdots+ \frac{1}{{n!\left( {n + 1} \right)!}}
 {{\left( {\frac{x}{2}} \right)}^{2n}} +  \cdots } \right] =  \\
 {\rm{  }} = \frac{1}{{{{\left( {N!} \right)}^2}}}
 {\left( {\frac{x}{2}} \right)^{2N}}\frac{2}{x}{I_1}\left( x \right) =
 \frac{1}{{{{\left( {N!} \right)}^2}}}{\left( {\frac{x}{2}} \right)^{2N - 1}}{I_1}\left( x
 \right)\,.
 \end{multline}

\noindent Now for any real $x$ and $\left| \nu \right|<2$ we have

\begin{equation}
\label{54} {\varepsilon _N} \le \frac{{1 + \left| \nu \right|}}{{1
+ {{\nu}^2}}}\exp \left( {0.6449{{\nu}^2} + 0.2021F} \right)
  \frac{1}{{{{\left( {N!} \right)}^2}}}{\left(
{\frac{x}{2}} \right)^{2N + 1}}{I_1}\left( x \right)\,.
\end{equation}

\noindent Elimination of $\left| \nu \right|$ in the last inequality gives an estimation for any $x$ and $\left| \nu \right|<2$:

\begin{equation}\label{540}
{\varepsilon _N} \le 15\frac{1}{{{{\left( {N!}
\right)}^2}}}{\left( {\frac{x}{2}} \right)^{2N + 1}}{I_1}\left( x
\right). \nonumber
\end{equation}

\noindent For instance for  $x \le 2$ inequality (\ref{540}) reduces to

\begin{eqnarray}
\label{55} {\varepsilon _N} \le \frac{{24}}{{{{\left( {N!}
\right)}^2}}}.
\end{eqnarray}

\noindent The last inequality shows that for the functions $A\left( x \right), B\left( x \right), C\left( x \right), D\left( x \right)$
and for  $x \le 2$ and $\left| \nu \right| \le 2$  a computational error less then ${\varepsilon _8} < 1.5 \times {10^{ - 16}}$ occurs after the computation of only eight terms of the corresponding series.

\section{Conclusion}

\noindent A new algorithm for computing the real valued solutions of some special functions namely the Gamma function with pure imaginary argument, the Bessel (and modified Bessel) function of purely imaginary orders was systematically declared.  We have used these developed Bessel functions for the investigation and calculation of strain fields in semiconductor structures \cite{Fohtung09, Riotte09}. In the above applied case the values of the arguments of the Bessel functions lie in the vicinity of unity. The algorithm for their computations is shown to be rapidly converging and gives a very favorable accuracy for as few as eight iterations.

The algorithm described may give an alternate approach to solving the ground state of the screened coulombs potential problem. It is also promising in solving the inverse scattering problem which is mostly approached via the inverse Green's function. Further studies of the asymptotic behavior of these functions are still warranted as they play a great role in a wide class of boundary value problems and integral and differential problems of wave scattering.


\bibliography{Matyshev_fohtung_bessel}

\begin{thebibliography}{8}
\expandafter\ifx\csname natexlab\endcsname\relax\def\natexlab#1{#1}\fi
\expandafter\ifx\csname bibnamefont\endcsname\relax
  \def\bibnamefont#1{#1}\fi
\expandafter\ifx\csname bibfnamefont\endcsname\relax
  \def\bibfnamefont#1{#1}\fi
\expandafter\ifx\csname citenamefont\endcsname\relax
  \def\citenamefont#1{#1}\fi
\expandafter\ifx\csname url\endcsname\relax
  \def\url#1{\texttt{#1}}\fi
\expandafter\ifx\csname urlprefix\endcsname\relax\def\urlprefix{URL }\fi
\providecommand{\bibinfo}[2]{#2}
\providecommand{\eprint}[2][]{\url{#2}}

\bibitem[{\citenamefont{Watson}(1945)}]{Watson45}
\bibinfo{author}{\bibfnamefont{G.}~\bibnamefont{Watson}},
  \emph{\bibinfo{title}{A treatise on the theory of Bessel functions}}
  (\bibinfo{publisher}{Cambridge Mathematical Library}, \bibinfo{year}{1945}).

\bibitem[{\citenamefont{Matyshev}(2000)}]{Matyshev00}
\bibinfo{author}{\bibfnamefont{A.}~\bibnamefont{Matyshev}},
  \emph{\bibinfo{title}{Isotrajectory Corpuscular Optics}}
  (\bibinfo{publisher}{SPb, Science (in Russian)}, \bibinfo{year}{2000}).

\bibitem[{\citenamefont{{E. Fohtung et al}}(2009)}]{Fohtung09}
\bibinfo{author}{\bibnamefont{{E. Fohtung et al}}}, \bibinfo{journal}{in
  preparation}  (\bibinfo{year}{2009}).

\bibitem[{\citenamefont{Lommel}(1871)}]{Lommel1871}
\bibinfo{author}{\bibfnamefont{E.}~\bibnamefont{Lommel}},
  \bibinfo{journal}{Math. Ann.} \textbf{\bibinfo{volume}{3}},
  \bibinfo{pages}{475} (\bibinfo{year}{1871}).

\bibitem[{\citenamefont{Bocher}(1892)}]{Bocher1892}
\bibinfo{author}{\bibfnamefont{M.}~\bibnamefont{Bocher}},
  \bibinfo{journal}{Annals of Math.} \textbf{\bibinfo{volume}{6}},
  \bibinfo{pages}{137} (\bibinfo{year}{1892}).

\bibitem[{\citenamefont{Abramovitch}(1948)}]{Greenberg48}
\bibinfo{author}{\bibfnamefont{G.~G.} \bibnamefont{Abramovitch}},
  \emph{\bibinfo{title}{Selected Problems of Mathematical Theory of
  Electromagnetic Phenomena}} (\bibinfo{publisher}{Acad. of Sciences Pub.,
  Moscow-Leningrad (in Russian)}, \bibinfo{year}{1948}).

\bibitem[{\citenamefont{Boole}(1844)}]{Boole1844}
\bibinfo{author}{\bibfnamefont{G.}~\bibnamefont{Boole}},
  \bibinfo{journal}{Phil. Trans. of the Roy. Soc. of London}
  \textbf{\bibinfo{volume}{134, P2}}, \bibinfo{pages}{225}
  (\bibinfo{year}{1844}).

\bibitem[{\citenamefont{{M. Riotte, E. Fohtung, A.A. Minkevich, et
  al}}({2009})}]{Riotte09}
\bibinfo{author}{\bibnamefont{{M. Riotte, E. Fohtung, A.A. Minkevich, et al}}},
  \bibinfo{journal}{{in preparation}}  (\bibinfo{year}{{2009}}).

\end{thebibliography}

\end{document}